# Synthesis, Crystal Growth and Epitaxial Layer Deposition of FeSe$_{0.88}$ Superconductor and Other Poison Materials by Use of High Gas Pressure Trap System

O. Tkachenko[1,2], A. Morawski[1], A. J. Zaleski[2], P. Przyslupski[3], T. Dietl[3], R. Diduszko[3], A. Presz[1] and K. Werner-Malento[3]

[1] Institute of High Pressure Physics PAS, Sokolowska 29/37, 01-142 Warszawa, Poland
[2] Institute of Low Temperature and Structural Research PAS, Okolna 2, 50-422 Wrocław, Poland
[3] Institute of Physics, Polish Academy of Sciences, al. Lotnikow 32/46, 02-668 Warsszawa, Poland

**Abstract.** The FeSe samples in the form of polycrystals, single crystals and thin films have been prepared and characterized. The synthesized material has been hot isostatically pressed under pressure of 0.45 GPa of 5N purity argon with the use of the high gas pressure trap system (HGPTS). Thin films have been fabricated by the mixed procedures with the use of DC sputtering from various types of targets and processed employing the HGPTS. The used HGPTS assures a full separation of the active volume for synthesis or crystal growth of material and the inert gas medium. The obtained FeSe$_{0.88}$ samples have $T_c$ between 8 and 12 K. The samples have been characterized by SEM, EDX, XRD, magnetic susceptibility and resistivity measurements.

## 1 Introduction

The recent discovery of unconventional superconductivity in LaFeAsO$_{1-x}$F$_x$ family with $T_c$ = 26 K [1] was unexpected and gave a new impulse to investigation of iron-based quaternary oxypnictides, other alloys that contains Fe element and planar Fe based compounds. A lot number of work has been published in a short period of time. [2-4] Later some research groups described a superconductivity in arsenic-free PbO-type α-FeSe$_x$ (x=0.80 to 0.97) family compounds [5-7] and has introduced a new interesting type of superconductors with $T_c$ among 8 K. FeSe system also have been investigated earlier [8]. This compound has very similar planar crystal sublattice as the layered oxypnictides and is much easier to fabricate and also does not contain As. From this point of view this PbO-type compound is very interesting for investigations and gives us promises to be used in industrial applications, such as superconducting metal sheathed wires and tapes. Earlier Y. L. Chen et al. reported of fabricating of SmFeAsO$_{0.8}$F$_{0.2}$ tantalum wire with $T_c$ = 52.5 K and $J_c$ up to 2 x 10$^5$ A/cm$^2$ at 10 K. [9]

## 2 Experiment

Our samples were prepared with nominal composition FeSe$_{0.88}$. In our investigations we focused on three types of this compound samples – polycrystals, single crystals and thin films. High-purity powders of reduced iron (99.9%) and granulated selenium (99.99%) were mixed and milled manually in ceramic mortar in argon glove box. Then it was placed in special configuration alumina oxide ampoules. The set of ampoules have been placed in the high pressure furnace. This configuration is named High Gas Pressure Trap System (HGPTS) – fig. 1(a). By using of HGPTS at the conditions showed in fig. 1(b), (T = 580 $^0$C, P of 5N purity argon 0.45 GPa and time 25 hours) we able to obtain polycrystalline samples.

The single crystals have been grown by the same method in HGPTS at the similar conditions – fig. 1(c), but we made the preheating up 800 $^0$C and then temperature was slowly decreasing to 600$^o$C through 50 hours (P = 0.45 GPa). This allowed us to obtain single crystals with the dimensions up to 1 x 1 mm plate like.

Thin films have been obtained by putting inside the ampoule with Fe-Se initial mixture and with different substrates for thin films – LaAlO$_3$ and SrTiO$_3$. Some of this substrates were clean, another were preliminary deposited in vacuum by *in situ* and *ex situ* FeSe$_{0.88}$ compound by magnetron deposition method. Further ampoules with Fe-Se mixture and specially fixed inside substrates were also placed in high pressure furnace and have been processed by HGPTS method – fig. 1(d), at 900 $^0$C then decreasing for 45 hours with P = 0.45 GPa of Ar.

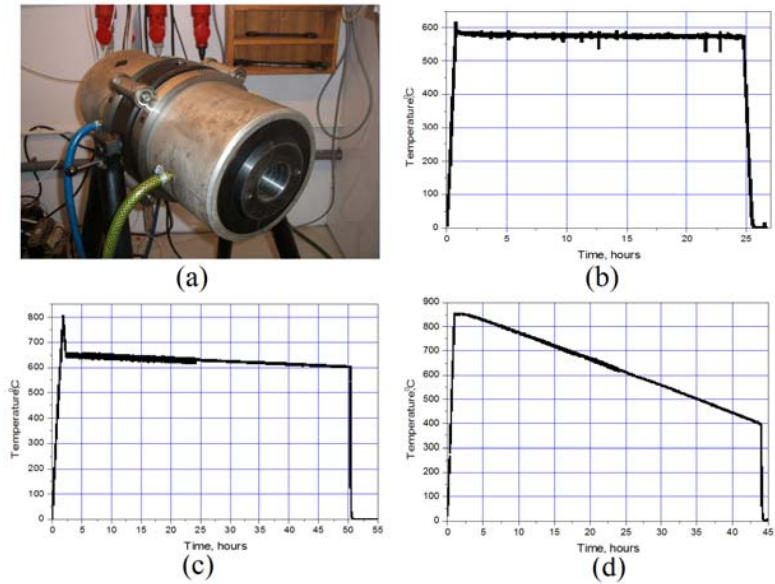

**Fig. 1.** (a) HGPTS system (up to 2000 $^0$C / 1.5 GPa). Typical annealing process in HGPTS for polycrytals (b), single crystals (c) and thin films (d)

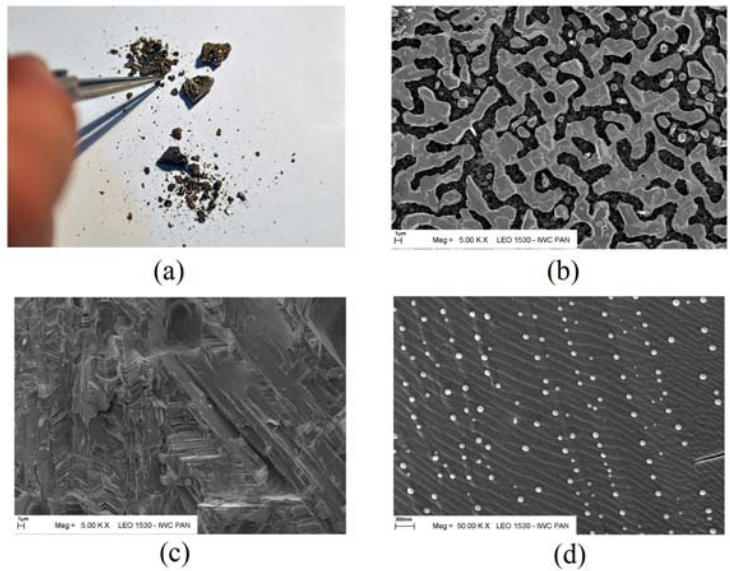

**Fig. 2.** (a) Optic picture of multiple FeSe$_{0.88}$ single crystals. (b) SEM. FeSe$_{0.88}$ thin layer from DC sputtering of *ex situ* material on SrTiO$_3$ substrate after high Se vapor pressure annealing. (c) SEM of the polycrystalline structure of FeSe$_{0.88}$. (d) Typical Se vapor condensation on the FeSe$_{0.88}$ layer.

## 3   Results and discussions

In fig. 2 one can see optical (a) and SEM pictures of our mono- and polycrystalline samples (c-d) and thin films (b).

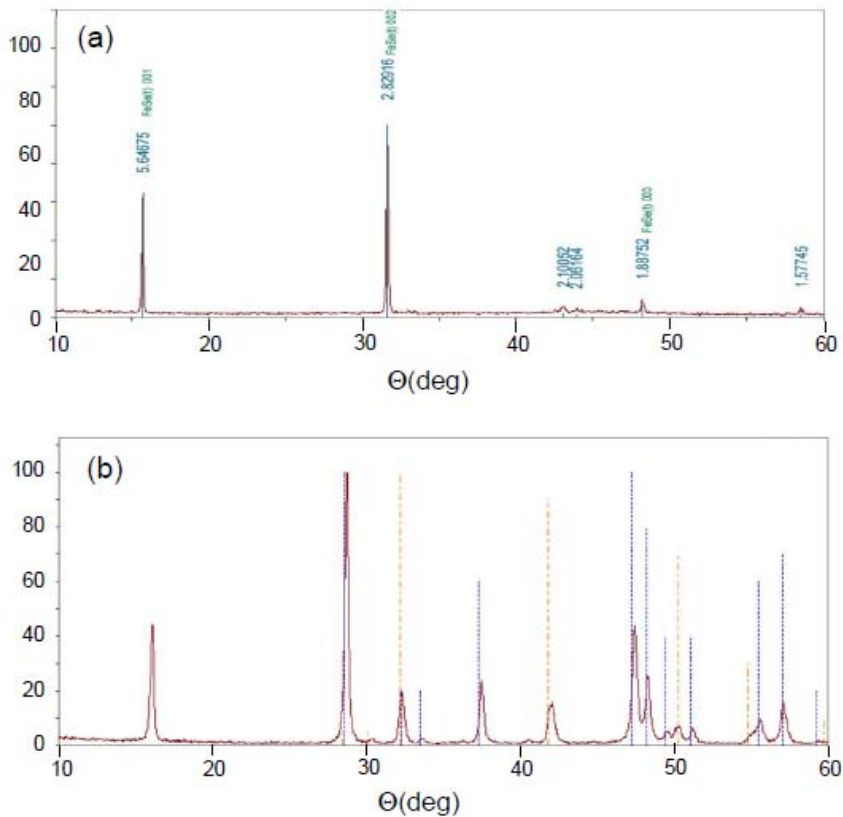

**Fig. 3.** XRD. (a) Thin layer of $FeSe_{0.88}$ from DC sputtering on $LaAlO_3$. The decrease of the *c* inter atomic distance is observed – suggesting the *a*, *b* directions large shrinking due to the applying substrate. High intensity of (001) reflex at 16 deg suggest the pure tetragonal phase. (b) Well crystallized polycrystalline sample of $FeSe_{0.88}$ – mainly tetragonal phase and only trace of hexagonal phase is seen.

The XRD analysis of thin film, obtained by high argon pressure annealing of the *in situ* deposited Fe+(0.88)Se material on $LaAlO_3$ substrate, shows the pure PbO type structure α-FeSe, characterized by slightly lower *c* parameter and sharp reflexes, see fig. 3(a), contrary to the *ex situ* deposited layers which posed rather normal *c* parameter and are usually dual phased with hexagonal phase which are very similar to the polycrystalline samples obtained at the same p, T, t, q (q – quench) conditions, see fig. 3(b).

The AC susceptibility measurements of the high gas pressure annealed initial powder show the polycrystalline structure which seems to be better crystalline at the first annealing process (see dual pick slop of the sample from 13.09.08 in fig. 4(c)) – the behavior correspond to the intra and inter grains proportion change of susceptibility, giving evidence for regular grains growth at this first annealing process. The milling of such material and next annealing at the same conditions results in higher granularity effect and lowering of the $T_c$ with characteristic higher $J_c$ (from Bean model estimations). The magnetically estimated $J_c$ depends strongly on the temperature and pressure conditions of phases formation and also by $T_c$ change due to the positive pressure effect of the $FeSe_{0.88}$ material.

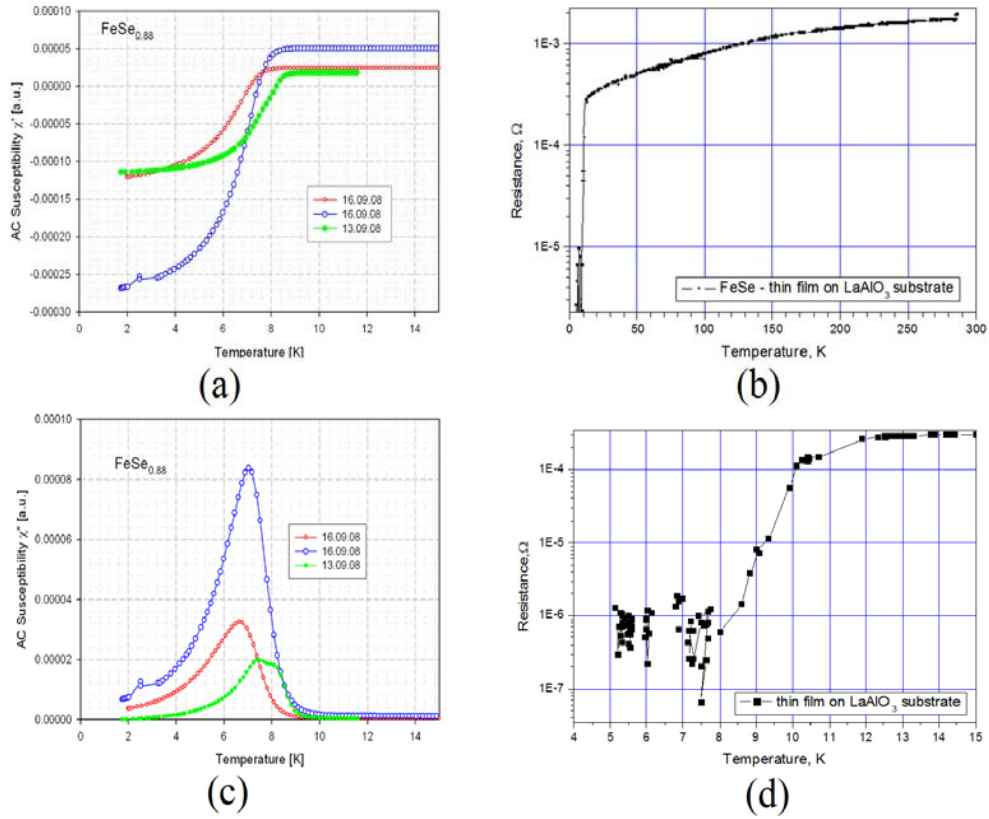

**Fig. 4.** (a), (c) Temperature dependence of AC susceptibility for 1kHz and 1 Gs for the differently obtained samples. Critical temperatures are similar and equal to about 10K. It is seen that because of the different amount of tetragonal phase present there is a difference in magnetic properties in normal state. The sample in which there is highest admixture of hexagonal phase clear paramagnetic susceptibility is observed above the $T_c$. (b), (d) The R=f(T) slope and superconducting transition of the thin layers of $FeSe_{0.88}$ on mono crystalline substrate $LaAlO_3$ obtained by HGPTS.

The $T_c$ values of such thin layers obtained at about 0.45 GPa argon pressure (as well as single crystals) posed $T_c$ of 11 K which is about 3 K higher then for ambient pressure obtained material, typically 8 K [10]. The results are in good agreement with the paper of Li et al. [5] where they found positive $dT_c/dp$ coefficient of about 9.1 K/GPa, the highest among all Fe-pnictide superconductors. Such behavior suggest not optimally doping of our material and can be related to the stress produced by use of high pressure on crystal growth, which in consequence provides the different order parameters of the Fe and Se atoms in the material. Such effect further produces a local anomaly which later shifts the phases transitions of the Fe-Se system to the higher temperature for samples obtained at higher pressures (to be reported later). So in our high pressure annealing we observed both temperature and pressure influences on the phases transitions relations, connected with superconductivity and magnetism coexistence but not evidently at the "true" the same crystallographic phase. The nature of these anomalies are still unknown.

## 4 Summary and conclusions

1. High Gas Pressure Trap System (HGPTS) has been developed and introduced, allow us to synthesize, anneal, carry out crystal growth and thin film formation of superconducting FeSe family compounds.
2. Good quality thin films and polycrystalline materials are obtained.
3. The single crystals can be easily extracted from the butch; their superconducting properties depend strongly on the local conditions of growing and quite frequently on the history of the phase transformations locally effected by application of p, T, t, q conditions.
4. The first sample of superconducting FeSe$_{0.88}$ wire was made.

**Acknowledgements.** One of us, O. Tkachenko, is acknowledging the support of NESPA grant under PR6 and Marie Curie Action programs. This work was partly supported by FunDMS Advanced Grant of European Research Council within the EU FP7 "Ideas" Program.